\newcommand{\um}{\ensuremath{\rm{\mu m}}}  
\newcommand{\sh}{\ensuremath{^{\rm{h}}}}   
\newcommand{\sm}{\ensuremath{^{\rm{m}}}}   
\newcommand{\galg}{G}   
\newcommand{\punit}{\ensuremath{\rm{10^{-10}\;W\,cm^{-2} \um^{-1} sr^{-1}}}}
\newcommand{\galb}{\emph{b}}  
\newcommand{\gall}{\emph{l}}  
\newcommand{\texttimes}{$\times$}  
\newcommand{\etal}{\emph{et al.}}
\newcommand{\bold}[1]{\textbf{#1}}
\newcommand{\err}{$1\sigma$}
\newcommand{\ISOtext}{
Based on observations with ISO, an ESA project with instruments funded by ESA
Member States (especially the PI countries: France, Germany, the Netherlands
and the United Kingdom) and with the participation of ISAS and NASA
}
\newcommand{\changed}[1]{#1}
\newcommand{\newchanged}[1]{#1}
\documentclass{aa}
\usepackage{txfonts}
\usepackage[final]{graphicx}	
\usepackage{subfig}
\usepackage{supertabular}
\begin{document}


   \title{Unidentified Infrared Bands 
          in the Interstellar Medium across the Galaxy\thanks{\ISOtext}}
   \author{J. Kahanp\"a\"a\inst{1},
	   K. Mattila\inst{1},
	   K. Lehtinen\inst{1},
           C. Leinert\inst{2}
\and	   D. Lemke\inst{2} 
		}
   \authorrunning{J. Kahanp\"a\"a \etal}
   \titlerunning{UIBs in the Interstellar Medium across the Galaxy}
   \offprints{J. Kahanp\"a\"a, \email{jere.kahanpaa@helsinki.fi}}

   \institute{Observatory, University of Helsinki, Po. Box 14,
		    FIN-00014 Helsingin yliopisto, Finland 
		\and    
              Max-Planck-Institut f\"ur Astronomie, K\"onigstuhl 17, 
                    D-69117 Heidelberg, Germany
		   }

   \date{Received December \hspace{1cm} 2002 / accepted \hspace{1cm} 2003 }


   \abstract{
      We present a set of 6-12 \um\ ISOPHOT-S spectra of the general 
      interstellar medium of the Milky Way. This part of the  spectrum  is
      dominated by a series of strong, wide emission features commonly called
      the Unidentified Infrared Bands.  The sampled area covers the inner Milky
      Way from \gall\ = -60\degr\  to +60\degr\ with a ten-degree step in
      longitude  and nominal latitudes \galb\ = 0\degr,$\pm$1\degr. For each
      grid position the actual  observed direction was selected from IRAS 100
      \um\ maps to  minimize contamination by point sources and molecular 
      clouds. All spectra were found to display the same spectral features.
      Band ratios are  independent of band strength and Galactic coordinates. A
      comparison of total observed flux in band features and IRAS 100 \um\
      emission, a tracer for large interstellar dust grains, shows high
      correlation at large as well as small (1\arcmin) scales. This implies a
      strong connection  between large dust grains and the elusive band
      carriers; the evolutionary history and heating energy  source of these
      populations must be strongly linked.  The average mid-infrared
      spectrum of the Milky Way is found to be \changed{similar to} the average
      spectrum of spiral galaxy \object{NGC 891} \changed{and the spectra of other spirals.} The common spectrum can therefore 
      be used as a template for the 6-12 \um\ emission of late-type spiral galaxies.      
      Finally, we show that interstellar extinction only
      weakly influences the observed features even at  $\lambda$~ = 10 \um,
      where the silicate absorption feature is strongest.
      }
\maketitle
%
%
%
%
\section{Introduction}
The 3--13 \um\ spectra of diffuse objects such as \ion{H}{II} regions, planetary and
reflection nebulae and Milky Way cirrus clouds  include a series of emission
bands collectively known as the unidentified infrared bands (UIR bands or
UIBs).  These structures have also been observed in external galaxies. 
Since the first detection of the 11.3 \um\ band by Gillett et al.
(\cite{Gillett73}) more than a dozen bands have been identified in various
astronomical spectra. Main bands always occur together, at 3.3, 6.2, 7.7, 8.6,
11.3 and 12.7 \um. Sometimes fainter companions at 3.4, 5.25, 5.65, 6.9, 9.7,
13.6, 14.1 and 16.5 \um\ are seen. The relative strengths of bands depend on
the observed object. This variation is probably related to changes in age and
chemical abundances in carrier molecules  since band ratios do not depend
strongly on strength or hardness of the local UV field (Chan et al.
\cite{Chan01}; Uchida et al. \cite{Uchida00}).

The exact nature of the carriers of the UIR bands is still unknown. 
It is generally agreed that the main bands are caused by bending and
stretching modes of carbon-carbon and carbon-hydrogen bonds in large organic
molecules; Holmlid (\cite{Holmlid00}) has, however, proposed an alternative
scenario based on de-excitation  of Rydberg matter.
All details of the chemical structure of UIB carriers are  still obscure. The
polyaromatic hydrocarbon (PAH) model, originally proposed by L\'eger \& Puget
(\cite{Leger84}) and Allamandola et al. (\cite{Allamandola85}) is frequently
used for analysis of the observed band widths and band ratios but other
proposed carriers like coal (Papoular et al. \cite{Papoular89}), hydrogenated
amorphous carbon (Duley \& Williams \cite{DW81}) and  quenched carbonaceous
composite (Sakata et al. \cite{Sakata84}) can not be ruled out at present.

In \cite{Puget85}, Puget et al. proposed that the IRAS 12 \um\ excess, first
reported by Boulanger et al. (\cite{Boulanger85}), is caused by a UIB component
in the diffuse Galactic emission. This hypothesis was supported by
detections of the 3.3 and 6.2 \um\ bands in the Galactic emission with
a balloon-borne instrument, AROME (Giard et al. \cite{Giard88}; Ristorcelli et al.
\cite{Ristorcelli94}) and later confirmed by detailed spectrophotometry by the
ISO and IRTS satellites (Mattila et al. \cite{Mattila96}; Onaka et al.
\cite{Onaka96}; Tanaka et al. \cite{Tanaka96}) and by the detection of UIB
emission from a single high-latitude cloud (Lemke et al. \cite{Lemke98}). It
has hence become clear that the UIB carriers are also common in the low-density
(0.01-100 \element[][]{H} atoms per cm$^{-3}$), low energy density regions known collectively
as the diffuse interstellar medium. Further support is provided by detection of
the interstellar 6.2 \um\ band in absorption in the IR spectra of Wolf-Rayet
stars (Schutte et al. \cite{Schutte98})




Most studies of properties of the UIBs and their carriers are based on
observations of high-density and high energy density environments such as
reflection or planetary nebulae, where the UIB carriers are expected to be
newly formed. The diffuse ISM provides us with a complementary set of
properties: there the UIB carriers are expected to be old and all volatile
species should have evaporated.  The low energy density limits the number of
possible carrier species. As equilibrium temperatures are not high enough to
produce these bands, transient heating of very small particles by single
photons must be considered (Greenberg \cite{Greenberg68}; Lemke et al.
\cite{Lemke98}; Boulanger et al. \cite{Boulanger98b}).

In this paper we present the results of an ISO guaranteed time project on the
distribution and properties of UIB carriers in the general interstellar matter
of the Galactic disk. Our aim is to answer the following questions: \\
1) What is the projected distribution of UIR band emission
   in the inner Milky Way? \\
2) Do the band ratios, band widths or scale height of the
   emitting layer change with Galactic longitude? \\
3) How does the UIB emission from the diffuse interstellar matter relate
   to other dust components (represented by IRAS 25, 60 and 100 \um\
   emission maps) and neutral or molecular gas? \\
\section{Observations and data reduction} \label{sec:obs_dr}
%
%
%
%
%
\begin{table*}[ht!]
  \caption[]{Summary of observed positions. Five leading 
  columns give the name and 
  equatorial and galactic coordinates for each position; 
  the remaining columns give the 
  UIR line intensities. The unit used is \punit. Statistical errors 
  for all intensities are also shown.}
  \[
  \begin{tabular}{l l l l l l l l l l l l l }
    \hline
    Name & RA(J2000) & Decl. & \gall (\degr )& \galb (\degr )& 
      I$_{6.2}$ & \err & I$_{7.7}$ & \err & I$_{8.6}$ & \err & I$_{11.3}$ & \err \\
    \hline
    \galg-60+05 &  12\sh13\sm11\fs 8&  -57\degr 19\arcmin 44\farcs0 & 297.77 & 5.16 
      & & & & & & & & \\
    \galg-60+01 &  12\sh 9\sm16\fs 6&  -61\degr 26\arcmin 48\farcs8 & 297.92 & 1.01 
      & 0.38 & $\pm$0.08 & 0.29 & $\pm$0.04 & 0.29 & $\pm$0.07 & 0.36 & $\pm$0.11\\
    \galg-60+00 &  12\sh 7\sm12\fs 0&  -62\degr 13\arcmin 39\farcs1 & 297.80 & -0.20 
      & 0.60 & 0.10 & 0.56 & 0.05 & 0.34 & 0.06 & 0.62 & 0.14 \\
    \vspace{0.09cm}
    \galg-60-05 &  11\sh59\sm6\fs 5&  -68\degr 49\arcmin 29\farcs8  & 298.21 & -6.43
      & & & & & & & & \\
    \galg-45+05$^2$ & 14\sh18\sm46\fs 9&  -55\degr 37\arcmin 9\farcs3  & 315.11 & 5.17 
      & & & & & & & & \\
    \galg-45+01 & 14\sh32\sm41\fs 8&  -59\degr 23\arcmin 39\farcs9 & 315.51 & 0.98  
      & 0.51 & $\pm$0.09 & 0.87 & $\pm$0.07 & 0.40 & $\pm$0.06 & 0.69 & $\pm$0.12 \\
    \galg-45+00 & 14\sh35\sm2\fs 7&  -60\degr 22\arcmin 28\farcs8  & 315.40 & -0.04 
      & 1.19 & 0.15 & 1.74 & 0.11 & 0.53 & 0.09 & 0.76 & 0.14 \\
    \galg-45-01 & 14\sh39\sm1\fs 6&  -61\degr 13\arcmin 40\farcs2  & 315.51 & -1.01 
      & 0.67 & 0.12 & 0.96 & 0.07 & 0.43 & 0.07 & 0.83 & 0.15\\
    \vspace{0.09cm}
    \galg-45-05 & 14\sh52\sm53\fs 8&  -64\degr 57\arcmin 40\farcs7 & 315.33 & -5.06 
      & & & & & & & & \\
    \galg-30+05 & 15\sh45\sm1\fs 8&  -48\degr 37\arcmin 23\farcs3   & 330.05 & 4.89   
      & & & & & & & & \\
    \galg-30+01 & 16\sh 0\sm39\fs 5&  -51\degr 35\arcmin 57\farcs0  & 330.10 & 1.00    
      & 1.14 & $\pm$0.19 & 1.45 & $\pm$0.12 & 0.53 & $\pm$0.11 & 1.17 & $\pm$0.17\\
    \galg-30+00 & 16\sh 3\sm38\fs 7&  -52\degr 18\arcmin 1\farcs5   & 329.99 & 0.17     
      & 2.20 & 0.28 & 2.83 & 0.19 & 0.89 & 0.16 & 2.07 & 0.23 \\
    \galg-30-01 & 16\sh 8\sm51\fs 2&  -53\degr  9\arcmin 46\farcs2  & 330.00 & -1.00  
      & 0.77 & 0.12 & 0.89 & 0.08 & 0.32 & 0.09 & 0.82 & 0.15 \\
    \vspace{0.09cm}
    \galg-30-05 & 16\sh27\sm22\fs 8&  -56\degr 11\arcmin 31\farcs5  & 329.79 & -5.03  
      & & & & & & & & \\
    \galg-15+05 & 16\sh43\sm43\fs 1&  -38\degr 14\arcmin 40\farcs9  & 344.93  & 5.01     
      & & & & & & & & \\
    \galg-15+01$^3$ & 17\sh 1\sm4\fs 8&  -40\degr 48\arcmin 43\farcs4  & 345.07 & 0.80  
      & 1.07 & $\pm$0.21 & 1.58 & $\pm$0.14 & 0.46 & $\pm$0.12 & 0.63 & $\pm$0.20  \\
    \galg-15+00 & 17\sh 5\sm11\fs 9&  -41\degr  7\arcmin 27\farcs5  & 345.29 & -0.01
      & 1.39 & 0.20 & 2.01 & 0.15 & 0.45 & 0.11 & 0.68 & 0.17  \\
    \galg-15-01 & 17\sh 7\sm51\fs 7&  -41\degr 58\arcmin 9\farcs6   & 344.92 & -0.91
      & 1.24 & 0.19 & 1.57 & 0.13 & 0.66 & 0.13 & 1.00 & 0.18  \\
    \vspace{0.09cm}
    \galg-15-05 & 17\sh27\sm31\fs 0&  -44\degr  7\arcmin 49\farcs4  & 345.22 & -6.43
      & & & & & & & & \\
    \galg-05+05$^1$& 17\sh14\sm53\fs 5&  -29\degr 43\arcmin 38\farcs3  & 355.65 & 5.16
      & & & & & & & & \\
    \galg-05+01 & 17\sh29\sm41\fs 3&  -32\degr 41\arcmin 11\farcs1  & 355.00 & 0.88
      & 1.01 & $\pm$0.23 & 1.55 & $\pm$0.17 & 0.46 & $\pm$0.15 & 1.36 & $\pm$0.24  \\
    \galg-05+00 & 17\sh32\sm53\fs 8&  -33\degr 10\arcmin 1\farcs4   & 354.97 & 0.05
      & 2.13 & 0.36 & 3.12 & 0.26 & 0.83 & 0.22 & 2.29 & 0.33  \\
    \galg-05-01 & 17\sh37\sm39\fs 1&  -33\degr 37\arcmin 39\farcs3  & 355.12 & -1.03
      & 0.58 & 0.16 & 0.90 & 0.12 & 0.22 & 0.13 & 0.55 & 0.20  \\
    \vspace{0.09cm}
    \galg-05-05 & 17\sh57\sm16\fs 9&  -34\degr 38\arcmin 47\farcs2  & 356.34 & -5.04
      & & & & & & & & \\
    \galg+05+05 & 17\sh35\sm11\fs 6&  -21\degr  1\arcmin 56\farcs7  & 5.47 & 4.99   
      & & & & & & & & \\
    \galg+05+01 & 17\sh47\sm2\fs 1&  -26\degr 21\arcmin 59\farcs7   & 2.36 & 1.07  
      & 0.73 & $\pm$0.21 & 0.86 & $\pm$0.11 & 0.25 & $\pm$0.13 & 0.52 & $\pm$0.20  \\
    \galg+05+00 & 17\sh53\sm52\fs 3&  -25\degr 52\arcmin 14\farcs7   & 3.57 & 0.00 
      & 1.73 & 0.28 & 2.77 & 0.21 & 0.77 & 0.19 & 1.81 & 0.26  \\
    \galg+05-01 & 17\sh56\sm30\fs 1&  -26\degr 53\arcmin 41\farcs7   & 2.98 & -1.02
      & 0.95 & 0.18 & 0.97 & 0.12 & 0.38 & 0.13 & 0.77 & 0.21  \\
    \vspace{0.09cm}
    \galg+05-05 & 18\sh14\sm32\fs 0&  -29\degr 18\arcmin 1\farcs4   & 2.82 & -5.67 
      & & & & & & & & \\
    \galg+15+05 & 17\sh50\sm52\fs 8&  -15\degr 11\arcmin 59\farcs8  & 12.43 & 6.01 
      & & & & & & & & \\
    \galg+15+01 & 18\sh15\sm28\fs 8&  -14\degr 47\arcmin 24\farcs6   & 15.71 & 1.03 
      & 1.29 & $\pm$0.19 & 1.32 & $\pm$0.13 & 0.65 & $\pm$0.13 & 1.31 & $\pm$0.21  \\
    \galg+15+00 & 18\sh19\sm15\fs 7&  -15\degr 16\arcmin 55\farcs8   & 15.70 & 0.00 
      & 3.17 & 0.41 & 4.05 & 0.28 & 1.51 & 0.22 & 2.90 & 0.32  \\
    \galg+15-01 & 18\sh23\sm17\fs 9&  -15\degr 30\arcmin 54\farcs6   & 15.96 & -0.97
      & 1.56 & 0.21 & 2.04 & 0.15 & 0.82 & 0.14 & 1.59 & 0.23  \\
    \vspace{0.09cm}
    \galg+15-05 & 18\sh46\sm16\fs 3&  -15\degr 56\arcmin 24\farcs2   & 18.11  & -6.08
      & & & & & & & & \\
    \galg+30+05$^4$& 18\sh27\sm26\fs6&    0\degr 45\arcmin 57\farcs3   & 29.51 & 4.99     
      & & & & & & & & \\
    \galg+30+01 & 18\sh40\sm30\fs 1&   -3\degr 29\arcmin 23\farcs4   & 28.58 & 0.84  
      & 0.99 & $\pm$0.17 & 1.30 & $\pm$0.11 & 0.57 & $\pm$0.11 & 0.81 & $\pm$0.14  \\
    \galg+30+00 & 18\sh42\sm8\fs 6&   -4\degr 32\arcmin 2\farcs1   & 27.84 & 0.00   
      & 2.29 & 0.29 & 2.91 & 0.18 & 0.86 & 0.14 & 1.65 & 0.20  \\
    \galg+30-01 & 18\sh45\sm53\fs 5&   -4\degr 55\arcmin 49\farcs1   & 27.91 & -1.02 
      & 0.64 & 0.14 & 0.87 & 0.09 & 0.48 & 0.09 & 0.50 & 0.14  \\
    \vspace{0.09cm}
    \galg+30-05 & 18\sh59\sm16\fs 4&   -7\degr 10\arcmin 46\farcs4  & 27.42 & -5.00
      & & & & & & & & \\
    \galg+45+05 & 18\sh51\sm54\fs 7&   11\degr 13\arcmin 19\farcs9  & 43.00 & 4.99     
      & & & & & & & & \\
    \galg+45+01 & 19\sh11\sm59\fs 7&   12\degr  4\arcmin 4\farcs7  & 46.00  & 1.00   
      & 0.25 & $\pm$0.07 & 0.41 & $\pm$0.06 & 0.33 & $\pm$0.06 & -0.05 & $\pm$0.08  \\
    \galg+45+00 & 19\sh13\sm34\fs 7&   10\degr 38\arcmin 58\farcs5  & 44.92 & -0.01   
      & 0.62 & 0.10 & 0.97 & 0.09 & 0.51 & 0.09 & 0.47 & 0.10  \\
    \galg+45-01 & 19\sh20\sm31\fs 0&   11\degr 39\arcmin 46\farcs8   & 46.61 & -1.04 
      & 0.35 & 0.08 & 0.53 & 0.06 & 0.27 & 0.06 & 0.18 & 0.10  \\
    \vspace{0.09cm}
    \galg+45-05 & 19\sh30\sm44\fs 2&    8\degr  2\arcmin 50\farcs7  & 44.61 & -4.96 
      & & & & & & & & \\
    \galg+60+05 & 19\sh27\sm45\fs 2&   27\degr 29\arcmin 10\farcs2   & 61.38 & 4.93    
      & & & & & & & & \\
    \galg+60+01 & 19\sh42\sm9\fs 5&   25\degr 16\arcmin 46\farcs5    & 61.01 & 1.04 
      & 0.38 & $\pm$0.07 & 0.24 & $\pm$0.05 & 0.17 & $\pm$0.05 & -0.20 & $\pm$0.07   \\
    \galg+60+00 & 19\sh45\sm49\fs 6&   24\degr 18\arcmin 18\farcs3   & 60.58 & -0.17 
      & 0.80 & 0.10 & 0.80 & 0.07 & 0.29 & 0.07 & 0.36 & 0.10  \\
    \galg+60-01 & 19\sh47\sm52\fs 6&   23\degr 22\arcmin 56\farcs0   & 60.02 & -1.04 
      & 0.40 & 0.06 & 0.37 & 0.04 & 0.08 & 0.04 & -0.02 & 0.07  \\
    \galg+60-05 & 20\sh 5\sm18\fs 8&   22\degr 22\arcmin 11\farcs3    & 61.2 & -5.00  
      & & & & & & & & \\
    \hline
    \end{tabular}
  \]
    1 = Very high dark current. \\
    2 = Has 7.7 \um\ 'ghost' band in the dark measurement. \\
    3 = Shows evidence of point source contamination in one of the four pixels.
        This pixel was omitted from the average value calculation. \\
    4 = Observed 7 months later than other \galg+30 points. \\
  \label{table:positions}
\end{table*}
\begin{figure}
  \centering
  \resizebox{\hsize}{!}{\includegraphics{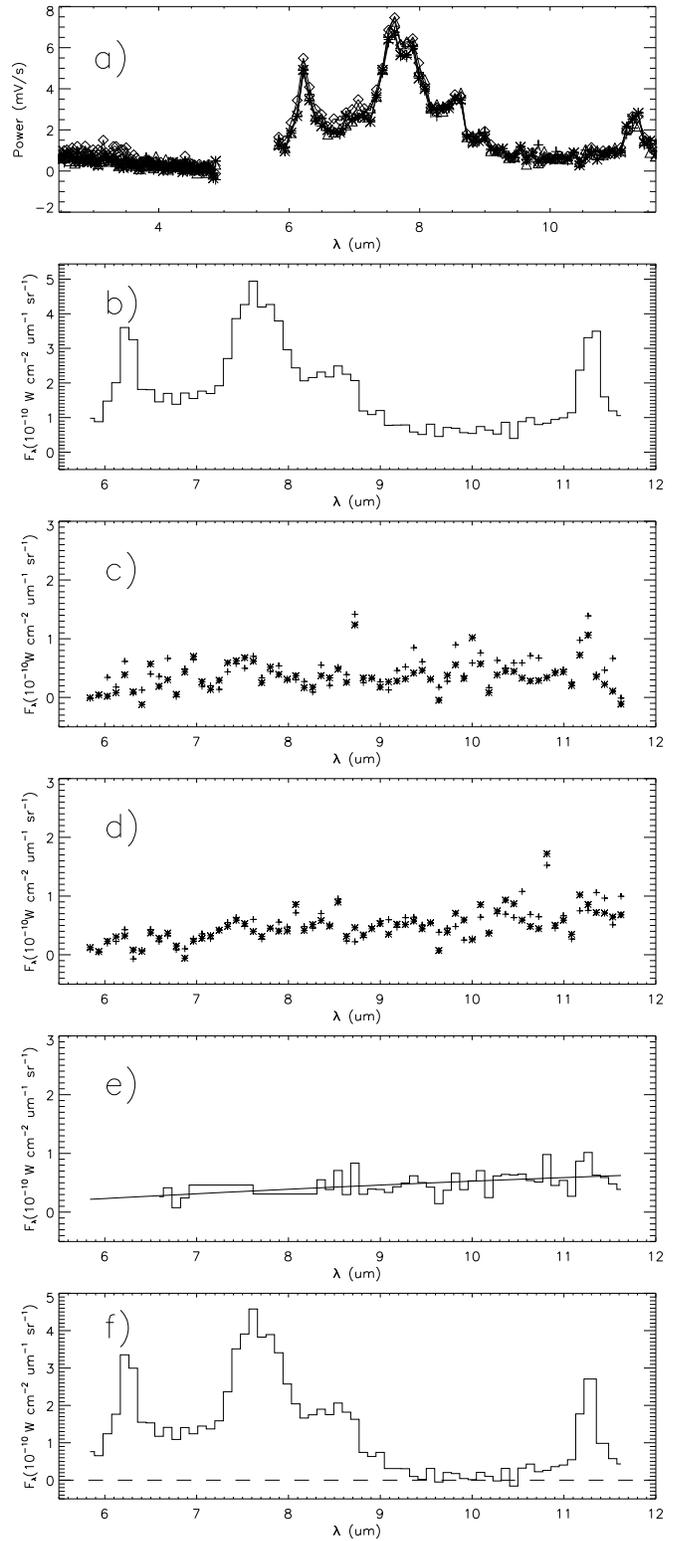}}
     \caption{
        Data reduction steps of the PHT-S spectrum of
	\galg+30+0. \bold{a)} Charge accumulation rate (mV/s) for the four individual 
	pointings. Data for all four pixels is shown. \bold{b)}
	Median of surface brightness (\punit). Zodiacal emission surface brightness at
	\bold{c)} \galb\ = $+5\degr$ and \bold{d)} \galb\ = $-5\degr$. \bold{e)} Average the of the zodiacal
	spectrum  at \galb\ = $\pm$5\degr\ and a polynomial fit to the data.
	\bold{f)}
	Final foreground-subtracted spectrum.}
  \label{fig:reduction_steps}
\end{figure}
\begin{figure}
  \resizebox{\hsize}{!}{\includegraphics{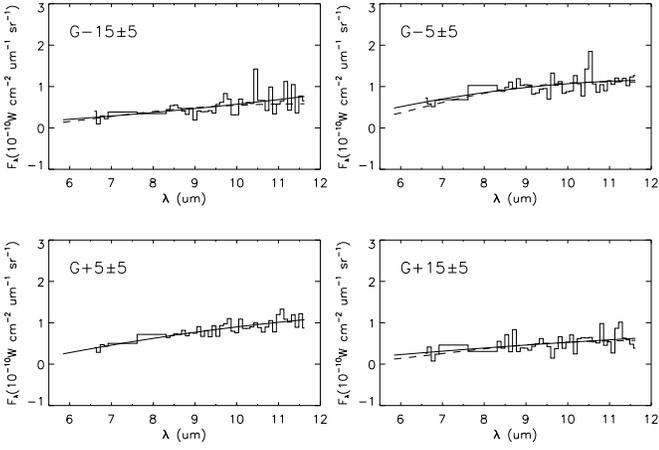}}
  \caption{
	 Zodiacal foreground emission for selected OFF-positions.
         A 2nd degree polynomial fit to the data is shown with a solid line and
         a model prediction of zodi\_emit (see Sect.~\ref{sec:obs_dr}) 
	 with a dashed line. In the 
	 case of \galg+5$\pm$5 the two lines completely coincide.
	 }
  \label{fig:zodiacal_light}
\end{figure}
%
The dataset consists of 49 small raster maps made with the ISOPHOT-S
spectrometer (Lemke et al. \cite{Lemke96}) onboard the Infrared Space
Observatory (ISO) (Kessler et al. \cite{Kessler96}). This low-resolution (R
$\approx$ 60, $\Delta\lambda \approx 0.1$ \um) instrument covers the wavelength
range  $\lambda $ = \mbox{2.5--11.6 \um} with a small gap at \mbox{4.9--5.8
\um.} Observations spanned a large part of the lifetime of ISO;  the first
dataset was recorded during revolution 81 (2/1996) and the last during
revolution 840 (3/1998). \changed{The equatorial and galactic coordinates of
the observed positions are  listed in Table~\ref{table:positions}.}
ISO Target Dedicated Time (TDT) numbers for each observation are listed in
Table~2\changed{, available only in electronic form at the CDS}.
\addtocounter{table}{1}
The observed fields sampled the inner Milky Way at  $\gall \approx \pm 5\degr,
\pm 15\degr, \pm 30\degr, \pm 45\degr$ and $\pm 60\degr$. For each nominal
longitude, five latitudes were observed:  two reference
measurements (OFF-positions) at $\galb \approx +5\degr$ and -5\degr\  
in order to check the zodiacal emission 
and three
ON-positions at $\galb \approx 0, \pm 1\degr$.  A single position included in
the original grid, \galg-60-1, was not  observed due to observing time
constraints. 

All observations at a given Galactic longitude were done within the same ISO
revolution to minimize variations in the zodiacal emission foreground due to
changes in solar aspect angle. The
contribution of Galactic emission in the  OFF-positions was minimized by
choosing the darkest positions in IRAS  12 and 100 \um\ maps close to the
nominal (\gall, \galb) positions. For the ON-positions the IRAS maps were
checked for point sources and the exact positions were then selected in regions
with no known Galactic or extragalactic IR sources.  CO maps were also 
consulted and regions with major molecular gas complexes were avoided.   The
resulting positions should be a reasonable sample of the general ISM:
emission from the tenuous diffuse ISM dominates even if regions with molecular
gas can never be completely avoided in this kind of sampling.


Each observation consists of a small raster map made with ISO Astronomical 
Observing Template P40 (see Laureijs et al. \cite{Laureijs00}); a standard 
32-second dark current and memory effect checking integration was followed  by
64-second sky measurements arranged in a small raster map. For the
ON-positions the map  size was $2 \times 2$ pixels, for OFF-positions $2 \times
1$ pixels. The distance between raster points was equal to the PHT-S aperture
size ($24\arcsec \times\ 24\arcsec$).


Data reduction was done with the ISOPHOT Interactive Analysis (PIA) program
Version 9.0.1.

A few observations were first processed manually to find the optimal reduction
procedure and the PIA batch processing mode was then used to create a
homogeneous set of calibrated spectra. Fig.~\ref{fig:reduction_steps} 
illustrates the reduction procedure. The following reduction steps were applied:
\begin{enumerate}
\item 
  2-threshhold deglitching (ie. detecting and removing cosmic ray
  events from raw data).
\item 
  Fitting ramps. Using the ramps subdivision technique tripled the number of
  usable data points and made the next step feasible.
\item 
  Removal of deviating points caused by 
  detector glitches.
  The first 15 seconds of each measurement were always discarded
  as values in this range are strongly affected by drift of the detector.
\item 
  Subtraction of dark current. We used the default orbital position-dependent
  dark current model as the dark measurement in the P40 observing mode  is
  dominated by memory effects. The top panel (a) in Fig.~\ref{fig:reduction_steps} shows the
  resulting spectrum for each of 
  the four pixels in the \galg+30+0 $2 \times 2$ raster map.
\item 
  Calibration from instrumental (V/s) to physical units ($\punit$). The default
  calibration scheme in PIA 9.0.1 was used. The P40 observing  mode is well
  suited for use of the new ISO dynamic calibration method, but since all
  observed fields are very faint, the resulting calibration was practically
  independent of the chosen calibration method. 
\item 
  Each 2\texttimes 2 raster map was checked for pixels with systematically
  higher values in either the short- or the long-wavelength part of the
  spectrum. If such a discrepancy was found the deviating pixel was rejected
  from further analysis. Only 
  one position (\galg-15+1) had signs of point source contamination in one of 
  the four pixels.
\item  
  The final search point in the (\gall,\galb )-grid was obtained by
  averaging over all good pixels in the raster maps. The second panel (b) in
  Fig.~1 presents the calibrated spectrum for \galg+30+0 in physical units
  (\punit).
\item 
  The contribution from zodiacal dust was estimated from the OFF-position
  spectra at $\galb = \pm 5\degr$. We fitted a 2nd degree polynomial to the 
  foreground emission measurements 
  and subtracted the resulting smooth approximation of the zodiacal emission
  spectrum from the ON-position spectra. Panels c and d  in Fig.~\ref{fig:reduction_steps}
  show the individual spectra at \galb\ = 30\degr, \gall\ = +5\degr\ and
  -5\degr, respectively. The next panel (e) shows the average together with
  the polynomial fit and the last panel (f) displays the final, calibrated and
  foreground-subtracted mid-IR spectrum at position \galg+30+0.
\end{enumerate}

\changed{
The error estimate for our data includes several components.  Error bars in
Fig.~\ref{fig:spectra} show the statistical errors in each spectrum.  The
uncertainty of the continuum level is mostly affected  by the accuracy of the
zodiacal foreground subtraction, which we estimate  to be better than 10\% of the
foreground level or less than 0.15 \punit.  Observed zodiacal light
levels at the OFF-positions were compared with model predictions by the
\textsf{zodi\_emit} method in the ISO Spectral Analysis Package (ISAP).  The
excellent match of both intensity and shape of the observed foreground spectra and
the models is shown in Fig.~\ref{fig:zodiacal_light}. The absolute calibration
accuracy of ISOPHOT-S is known to be around 10\% and relative accuracy of feature
intensity within the spectrum is 20\% (Laureijs et al. \cite{Laureijs00}) Finally,
it should be noted that the random noise pattern is slightly, but definitely
non-gaussian with a preference for high values. This effect is probably caused by
only partial correction for minor cosmic ray hits during  exposures.}

%
A few spectra are also affected by one to three  systematically deviating
detector pixels; these pixels were identified by the  very high dark current
and ignored in the analysis.

According to Tanaka et al. (\cite{Tanaka96}) and Giard et al. (\cite{Giard89})
the  3.3 \um\ band peak intensity in the plane of the Milky Way is 0.1--1~
\texttimes~ \punit -- less than the 1-$\sigma$ noise level of our spectra at
3.3 \um. No signs of the 3.3 \um\ peak were found in the short-wavelength part.
The short-wavelength parts of the spectra were not analyzed further.
%
%
\section{Results}
\begin{subfigures}
\begin{figure*}
  \centering
  \resizebox{\hsize}{!}{\includegraphics{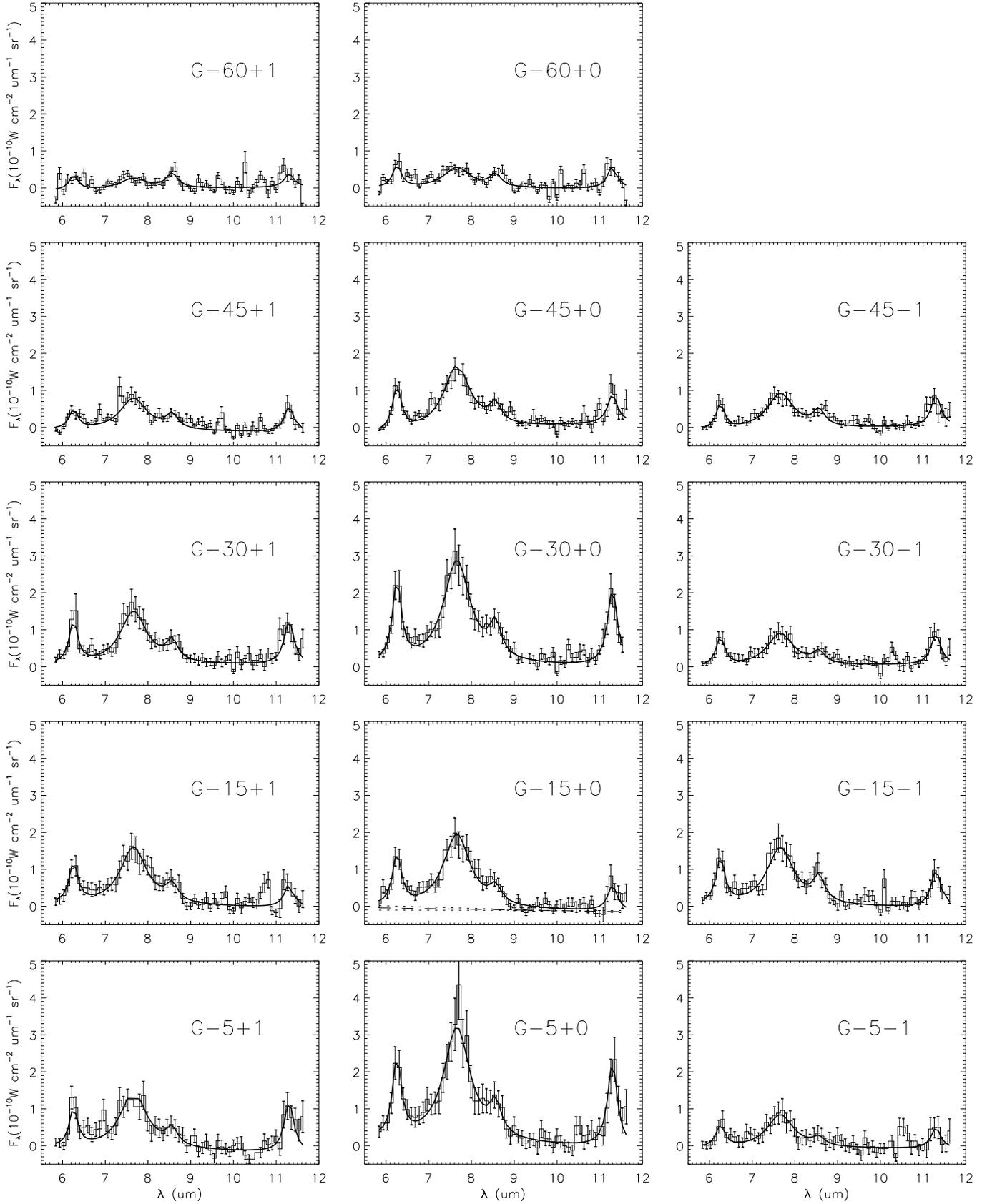}}
  \caption{
     UIB spectra of the general interstellar matter. The southern quarter 
     of the inner Milky Way (\gall = -60\degr\ -- -5\degr). For each position the 
     observed spectrum is shown as a histogram line and a fit 
     using four Cauchy profiles and a linear continuum as a 
     continuous line. For G-15+0 the linear component \changed{with 1-$\sigma$ 
     error limits} is also 
     displayed separately as a dashed line.}
%
  \label{fig:spectra1}
\end{figure*}
\begin{figure*}
  \resizebox{\hsize}{!}{\includegraphics{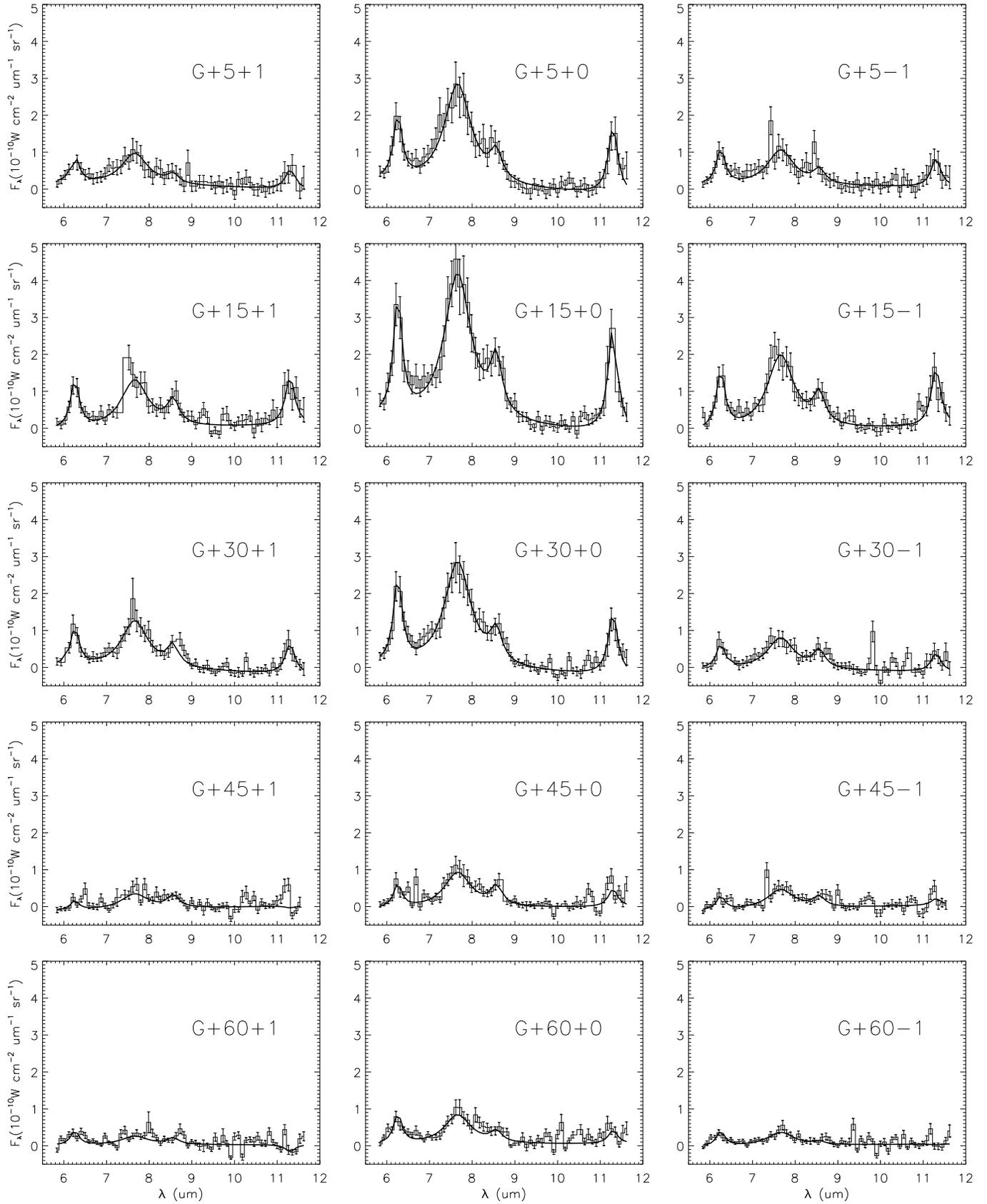}}
  \caption{
     UIB spectra of the general interstellar matter. The northern quarter 
     of the inner Milky Way (\gall = +5\degr\ -- +60\degr). 
     Data and fit as in Fig.~\ref{fig:spectra1}.}
  \label{fig:spectra2}
\end{figure*}
\label{fig:spectra}
\end{subfigures}
\clearpage
Visual inspection of the resulting spectra presented in Fig.~\ref{fig:spectra}
reveals a number of common properties:
\begin{itemize}
\item The 6.2, 7.7 and 11.3 \um\ UIR bands are widespread in the diffuse ISM. 
  All observed fields within $\gall < 45\degr$ display these bands.
  At \galg+60+0 and \galg-60+0 the bands are clearly visible, but the off-plane
  positions at these longitudes display only traces of UIR bands.
\item  
  The flux level of all bands at $\galb = \pm1\degr$ is roughly one half of the
  flux at $\galb = 0\degr$.  If this relation also holds for the outer
  longitudes the lack of a reliable detection of band structures at
  \galg$\pm$60$\pm$1 is caused by the limited sensitivity of the spectrometer. 
\item   
  The 8.6 \um\ band is visible at most positions. Spectra with high S/N ratio
  clearly shows this feature as a well-defined bump on the shoulder of the
  broad 7.7 \um\ band. No fainter bands are detected.
\item The spectra are similar to those observed in molecular clouds
  (Boulanger et al. \cite{Boulanger96}; Cesarsky et al. \cite{Cesarsky96b}),
  reflection nebulae (Cesarsky et al. \cite{Cesarsky96a}), spiral 
  (Mattila et al.~\cite{Mattila99}) and starburst galaxies (Lutz et al.~\cite{Lutz99}).
%
%
\item 
  The widths, shapes, relative strengths and positions of the main bands are 
  seemingly uncorrelated with Galactic position or with the total UIR band
  flux.
\item 
  No position displays the clear continuum seen in hot sources such as the
  \object{M17} \ion{H}{II} region (see Cesarsky et al. \cite{Cesarsky96a}). Flux levels
  in the 9--11 \um\ range are very low in all spectra.  The average spectrum in
  Fig.~\ref{fig:faintlines} sets the upper limit of continuum emission near 10
  \um\ to $< $10\% of the 7.7 \um\ band peak flux. The contribution from
  stellar atmospheres, which is clearly seen in the short-wavelength (2.5--4.9
  \um) part of the spectrum (Fig.~\ref{fig:reduction_steps}a), is negligible in
  the long-wavelength part of the spectrum.
\item 
  Position \galg-15+0 displays seemingly aberrant behavior: the flux levels
  at the Galactic plane are only slightly higher than those at \galb\ =
  $\pm$1\degr\  while all other longitudes behave in a consistent  fashion.
  Closer examination of the IRAS 100 \um\ map reveals a very dark spot exactly
  at the observed position. 
  It seems  clear that we have in this case inadvertently
  selected a line-of-sight with little interstellar matter.
\end{itemize}
\subsection{Band widths and central wavelengths}
The UIR bands are frequently approximated and modelled with  Cauchy profiles:
\begin{equation}
F(\lambda) = \frac{h}
                  {1 + 4(\frac{\lambda-\lambda_0}{\sigma})^2},
\end{equation}
where $h$ is the height of the band and $\sigma$ is the full width at half
maximum (FWHM), superimposed on a linear continuum component  (see Boulanger et
al. \cite{Boulanger98a}). We adopted this method and fitted all ON-position
spectra with a combination of four Cauchy profiles and a linear component using
the \verb+curvefit+ function in IDL. \changed{No significant variability in the 
central wavelengths or band widths as a function of band strength or Galactic position was
detected during the first round of modeling. These two parameters for each band were
then fixed to the average values (listed in
Table~\ref{table:band_positions_and_widths}).}
Widths are comparable to other published values (see Table 7 in Li \&
Draine (\cite{Li01}) and references therein); the sparse literature on band
widths in the diffuse ISM is summarized in  Table~\ref{table:observed_widths}.
These values are also in agreement with the quantitative PAH model by Li
\& Draine if instrumental broadening is taken into account.
The central wavelengths are somewhat shifted compared to values listed in
Mattila et al. (\cite{Mattila99}); this is not a true difference between NGC
891 and the Milky Way, but caused by a change in the official ISO\changed{PHOT-S} wavelength
calibration.
\changed{
Final UIR band modeling used only the height of each band and the 
slope and zero point of the linear background as free parameters, 
giving a total of six free parameters.   
}
Fig.~\ref{fig:spectra} shows the fits (solid lines) overplotted on original
data.

The infrared continuum component is minimal in all spectra. 
\changed{As an example, the continuum component for position G-15+0 
is shown in Fig.~\ref{fig:spectra1}. The dashed continuum line is 
flanked by the steepest and flattest continuum (dotted lines) allowed within 
1-$\sigma$ error limits. Note that this fit and the error margins
are all within 0.15 \punit\ of zero and thus within the estimated 
error of our foreground subtraction accuracy.}
Our foreground 
subtraction method would eliminate a Galactic continuum emission component 
from our spectra, if and only if such a continuum had a very broad latitudinal
distribution at 6--11 \um\ with essentially the same values at  $\galb =
\pm5\degr$ and at $\galb = 0\degr$.  No such Galactic component is seen in the
COBE-DIRBE or IRAS 12 \um\ all-sky maps.  It is therefore clear that no
significant, mid-IR continuum at 6--11 \um\ is present in the sampled area.

\begin{table}
   \[
      \begin{tabular}{l l l}
	 \hline
	 \hline
	 band name & center  ($\lambda_0$) & width ($\sigma$)\\
	  & \um   &   \um  \\
	 \hline
	 6.2 & 6.26    $\pm$0.01  & 0.24 	\\%
	 7.7 & 7.66    $\pm$0.02  & 0.76 	\\%
	 8.6 & 8.56              & 0.33 	\\
	 11.3 & 11.30   $\pm$0.02  & 0.27 	\\%
	 \hline
	 \end{tabular}
   \]
   \caption[]{UIR band widths and central wavelengths \changed{in the diffuse interstellar
   matter.}}
   \label{table:band_positions_and_widths}
\end{table}
%
\begin{table*}
   \[
\begin{tabular}{l l l l l l l l}
   \hline
   \hline
   Source & 6.2 (\um ) & 7.7 (\um) & 8.6 (\um) & 11.3 (\um) &
   Resolution (\um)& Instrument & Notes\\
   \hline
    Mattila et al. \cite{Mattila96} & 0.20  & 0.71 & 0.42 & 0.27 & 0.1 & ISO/PHT-S & emission\\
    Onaka et al. \cite{Onaka96} & 0.67  & 0.91 & 0.55 & 0.63 & 0.3 & IRTS/MIRS & emission\\
    Schutte et al. \cite{Schutte98} & 0.15 & - & - & - & 0.004 & ISO/SWS & absorption\\
    this study & 0.24  & 0.76 & 0.33 & 0.27 & 0.1 & ISO/PHT-S & emission\\
   \hline
   \end{tabular}
   \]
   \caption{Published UIR band widths (FWHM, in \um) in the diffuse interstellar 
              medium.}
   \label{table:observed_widths}
\end{table*}

\subsection{Band ratios and correlations}
\begin{figure}
  \resizebox{\hsize}{!}{\includegraphics{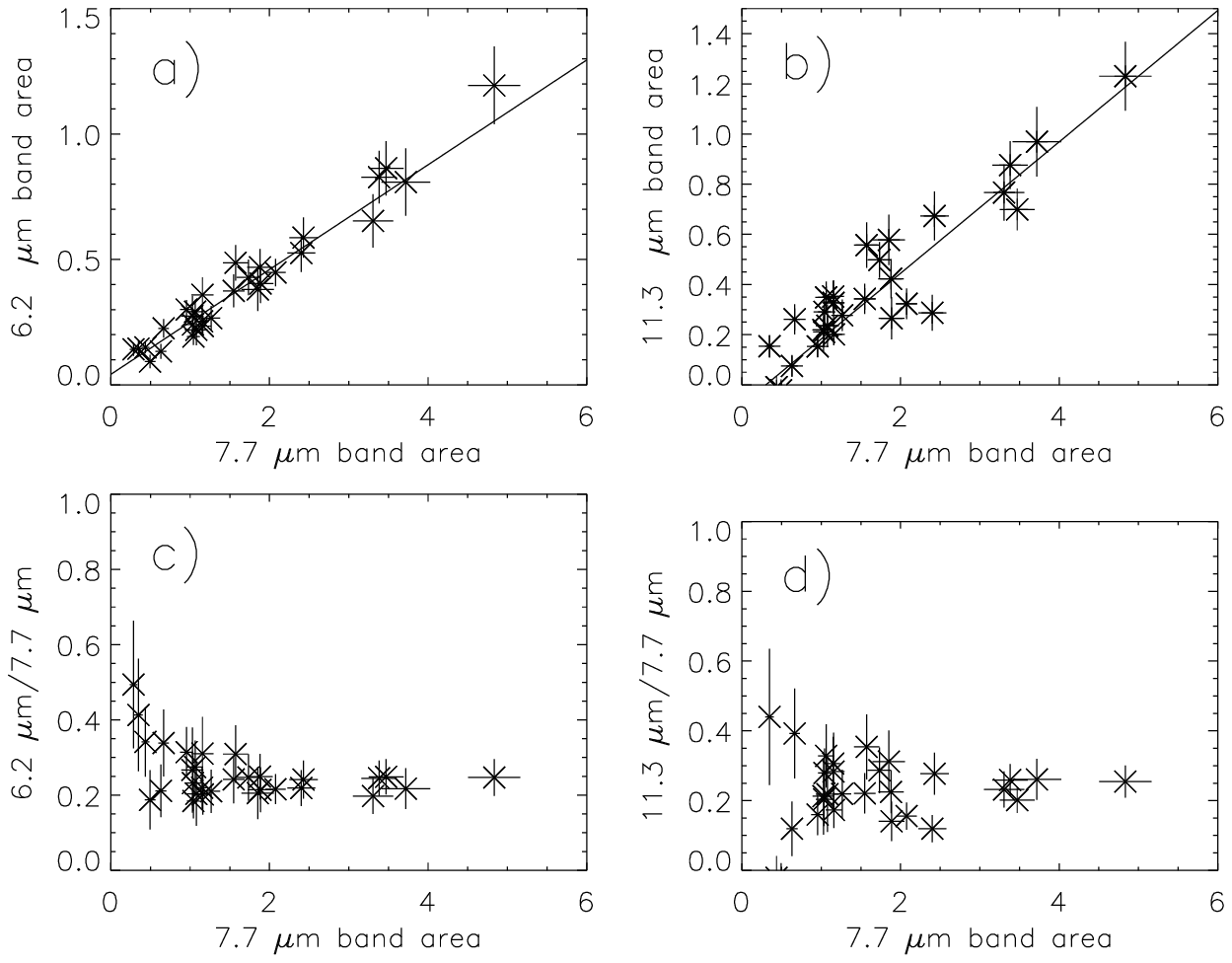}}
  \caption{       
	 \bold{a)} 6.2 \um\ vs. 7.7 \um\ band area (\punit).
	 \bold{b)} 11.3 \um\ vs. 7.7 \um\ band area.
	 \bold{c)} 6.2/7.7 \um\ band ratio vs. 7.7 \um\ band area.
	 \bold{d)} 11.3/7.7 \um\ band ratio vs. 7.7 \um\ band area. 
	 The solid lines in panels a) and b) show a 
	 least-square fit to the data. The resulting band 
	 ratios are 0.21 and 0.27 for the 6.2/7.7 \um\ and 11.3/7.7
	 \um\ pairs.
	 }
  \label{fig:correlation}
\end{figure}

\changed{We define the band area or band intensity as
the total area of the Cauchy profile, which can easily be calculated from 
band parameters: $A = 0.5 \pi h \sigma$. The total band area is then defined 
as the sum of band areas of the four bands seen in our ISOPHOT-S data.}
Any systematic changes in the relative band strengths with the total strength
of the UIR bands should be observable as displacements or nonlinearities in
Fig.~\ref{fig:correlation}, which displays the 6.2 and 11.3 \um\ band areas  as
a function of the 7.7 \um\ band area. No such structures can be seen in either
one of the band ratio plots. Least-squares fits  (shown in
Fig.~\ref{fig:correlation} as solid lines) to the band ratios yield average
ratios of 0.21 and 0.27 for the 6.2/7.7 \um\ and 11.3/7.7 \um\ band pairs,
respectively. The uncertainty for each ratio is $\pm$0.01. The standard
deviations of individual ratios are 0.07 and 0.13.

The two lower panels in Fig.~\ref{fig:correlation} show the behavior of band ratios
as a function of the 7.7 \um\ band strength for individual points.  11.3/7.7
\um\ flux ratios do not seem to have any systematic trends except the expected
decrease of noise towards higher intensities, while the 6.2/7.7 \um\ flux ratios
suggests that there may be a trend towards higher 6.2/7.7 \um\ ratio values at
small intensities.

%
\begin{figure}
  \resizebox{\hsize}{!}{\includegraphics{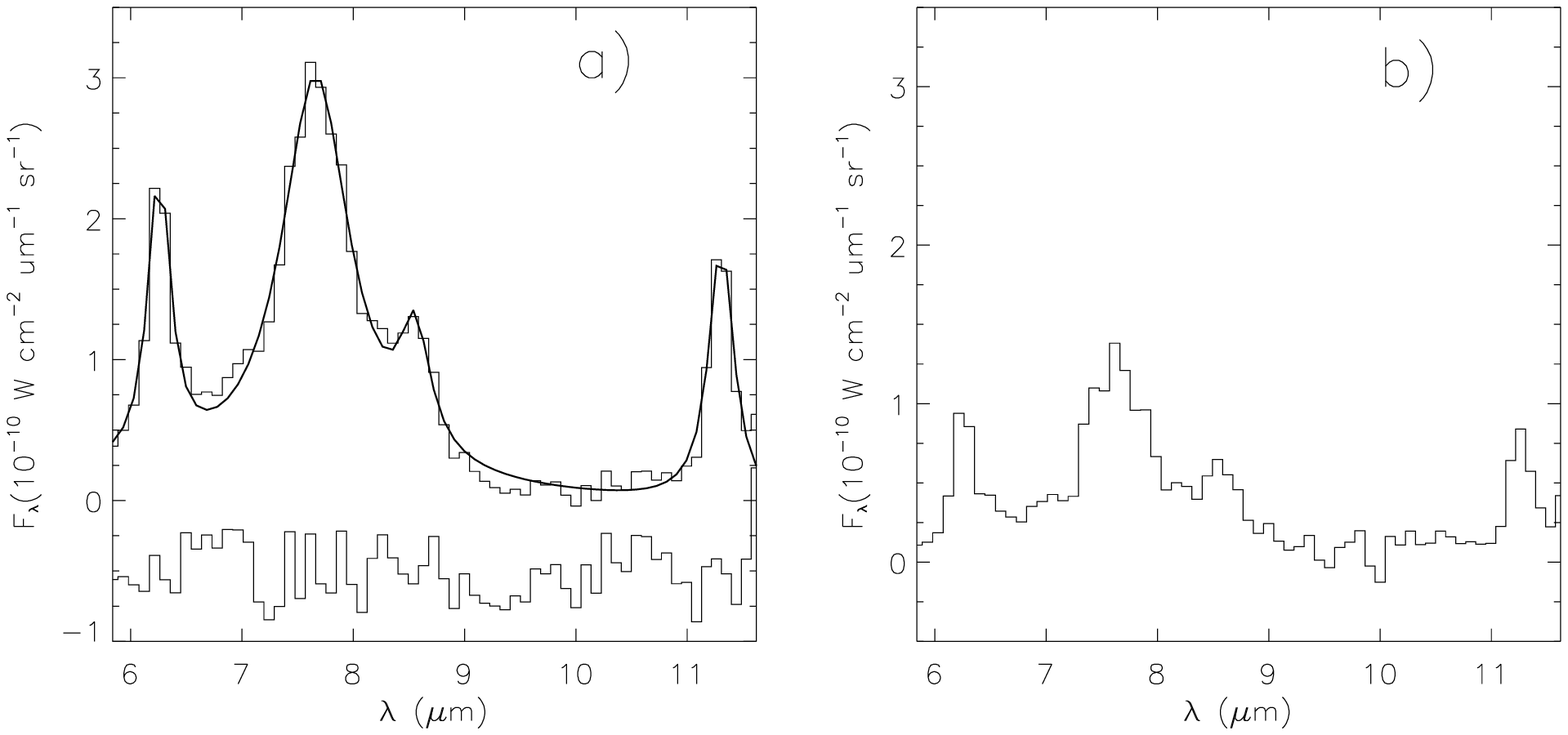}}
  \caption{
	 \bold{a)} The
	 average of six points with strongest emission 
	 (\galb\ = 0\degr, \gall\ = $\pm$5\degr, $\pm$15\degr and $\pm$30\degr) 
	 together with a
	 four-band fit to the average; the residual shown has been shifted 
	 downwards by -0.5 \punit\ and multiplied by a factor of 5 for clarity.
	 \bold{b)} The average over all spectra at 
	 Galactic latitude $\galb = \pm$ 1\degr.
	 }
  \label{fig:faintlines}
\end{figure}
%
%
\subsection{Weak UIR bands}
We searched for the fainter UIR bands at 6.9 and 9.7 \um\ in an average
spectrum encompassing the points between +30\degr\ and -30\degr\ 
Galactic longitude and \galb\ = 0\degr. The
resulting low-noise spectrum  (shown in
Fig.~\ref{fig:faintlines}) was fitted with the same  combination of four
Cauchy profiles and a linear continuum as the individual spectra.  The
residual does not reveal any clear structures; this sets the upper limit for
these bands at 0.3 \texttimes \punit, 
\changed{a value much larger than the upper limit calculated
by Onaka et al. (\cite{Onaka96}) from a larger data set covering 
some 100 square degrees around \gall = 50\degr, \galb=0\degr.}

Some excess is present  near 6.9 \um\
but it might also be caused by a non-Lorentzian shape of the 7.7 \um\ band. 
Any signal at 9.7 \um\ is masked by the noise. 
For comparison Fig.~\ref{fig:faintlines} also shows the average of all 
off-plane spectra (\galb\ = $\pm$1\degr); no difference in the band shapes
between on-plane and off-plane positions is seen.
\subsection{Distribution of UIB emission within the Milky Way}
\label{section:distribution}
%

\begin{figure}
  \centering
  \resizebox{0.92 \hsize}{!}{\includegraphics{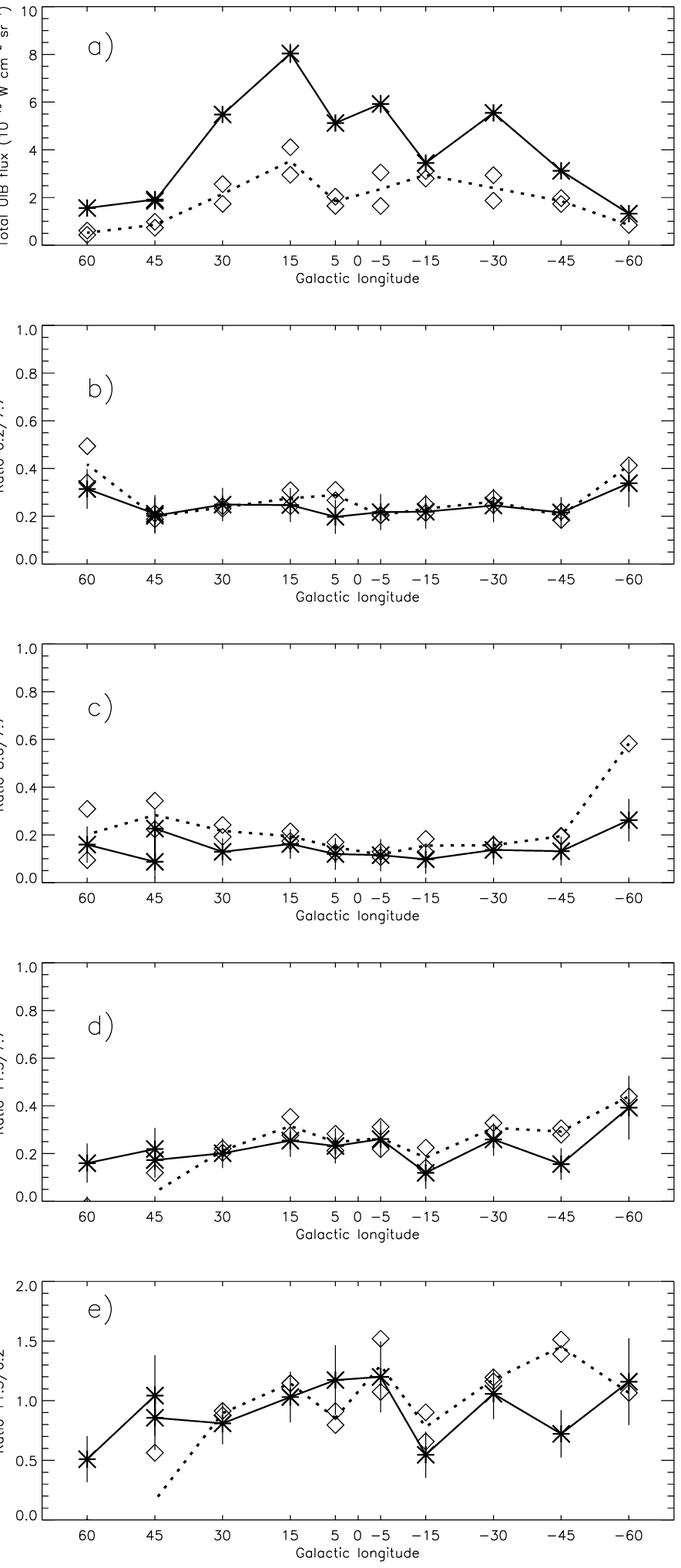}}
  \caption{
\changed{         Galactic distribution of UIB bands:
	 \bold{a)} Total area of the four main bands.}
	 \bold{b)} 6.2/7.7 \um\ band ratio.
	 \bold{c)} 8.6/7.7 \um\ band ratio.
	 \bold{d)} 11.3/7.7 \um\ band ratio.
	 \bold{e)} 11.3/6.2 \um\ band ratio.
	 In each panel stars and a line are used for  
	 values at the Galactic plane and 
	 triangles for
	 \galb\ = $\pm$1\degr. A dotted line traces the average of 
	 values at \galb\ = $\pm$1\degr. \changed{For clarity error bars are 
	 plotted only for Galactic plane values.}
	 }
  \label{Fig_map_ratios}
\end{figure}

The longitudinal distribution of the UIB emission is sampled at ten positions by
our dataset: at each longitude three latitudes (\galb\ = 0,$\pm$1\degr) were
probed. The resulting on-plane and off-plane longitude profiles of the
UIR bands and band ratios are shown in Fig.~\ref{Fig_map_ratios}. 
The on-plane profile
shows a strong concentration towards the central 90\degr\ of the Galactic plane
and strong variability within this region. The off-plane profile mimics the
on-plane values but displays less variation in the central region. No warp of
the diffuse dust layer is seen; the behavior of the \galb\ = +1\degr\ and
\galb\ = -1\degr\ profiles is identical in the outer (\gall\ $>$ 30\degr) part
of the sampled area. However, there is a clear systematic asymmetry
between the outer parts of the Galaxy: both on-plane and off-plane emission  is
weaker at \gall\ = +45\degr\ -- +60\degr\ than at \gall\ = -45\degr\ --
-60\degr. This asymmetry is also seen  in the large dust grain distribution
(COBE-DIRBE 240 \um\ data) and stellar  radiation (see Drimmel \& Spergel
\cite{Drimmel01} and references therein).

The variation \newchanged{of band ratios} 
seen in Fig.~\ref{Fig_map_ratios} does not deviate from a noise pattern 
since the statistical errors for individual band ratios are close to the 
standard deviation of the distribution of the ratios and no clear longitudal 
pattern is seen.

%
%

In order to estimate the angular height of the UIB-emitting layer at each
sampled Galactic longitude we fitted a Gaussian profile to the three observed
points at each nominal longitude. Due to the limited number of
observations, the FWHM and the peak flux at the Galactic plane were the only
free parameters while the center of the distribution was assumed to be on the
Galactic plane. 
No statistically significant trends were found in the longitudinal distribution 
of FHWM values except the high FWHM value at \gall\ = -15\degr\ caused by the
anomalously low flux at \galg-15+0. 
Our average FWHM value for the UIB component, 1.49\degr\ $\pm$
0.08\degr\ is close to the FWHM value (1.6\degr) found by Ristorcelli
et al. (\cite{Ristorcelli94}) for the 6.2 \um\ band in the inner Milky Way and 
the IRAS 12 \um\ FWHM value (1.5$\pm$0.1\degr).
The deviating longitude \gall\ = -15\degr\ was excluded from the average.

The latitude distribution of 3.3 \um\ band emission has two components: 
a narrow one with a FWHM of 1.6\degr\ and
a wide one with a 9-degree FHWM (Giard et al.
\cite{Giard89}). The present observations do not allow us to
exclude the possibility of a wide component in UIR bands between 
5.8 and 11.6 \um. The 
OFF-position measurements at $\pm$5\degr\ provide only an upper limit
($<$0.1 \punit)  for the flux levels at higher latitudes.


%
%
%
\section{Discussion}
The diffuse ISM provides us with a test laboratory where the properties  of all
interstellar components are uniform. Due to the high  age of the diffuse
component, we expect to see little large-scale variation  in the properties of
the UIB emitters and this is indeed what is observed.  Variations caused by
evolution of dust grains and strong UV fields have been diluted away. Another
stabilising effect is the independence of shape of UIBs of the  intensity of
the heating UV field. This is a general property of all models that rely on
single-photon heating; it holds as long as the UV flux does not ionize or
otherwise permanently modify the carrier particles.  These properties combine
to form a universal template spectrum of old  UIB carrier populations in cool
environments, which is observed over and  over again in the diffuse ISM of our
own Galaxy and in other quiescent  spiral galaxies.  Unfortunately, the general
insensitivity of UIB carriers to nuances of  ISM conditions make the bands
rather useless as probes of physical conditions  of the diffuse interstellar
matter. 

Because of the diffuse, but still inhomogeneous nature of the  general ISM, each
one of our observations samples a range of environments. A typical
line-of-sight will pass through several small  molecular  clouds in addition to
the tenuous intercloud medium.  The very existence of a template spectrum --
ie. a spectrum  that matches all positions -- proves that cloud-to-cloud
variation of  the spectral shape must be small. This has already been pointed
out by  Chan et al. (\cite{Chan01}), who analysed a larger sample of observations
which did not specifically target the diffuse component.  





%

\subsection{Correlation with other ISM components}
\begin{figure}
  \resizebox{\hsize}{!}{\includegraphics{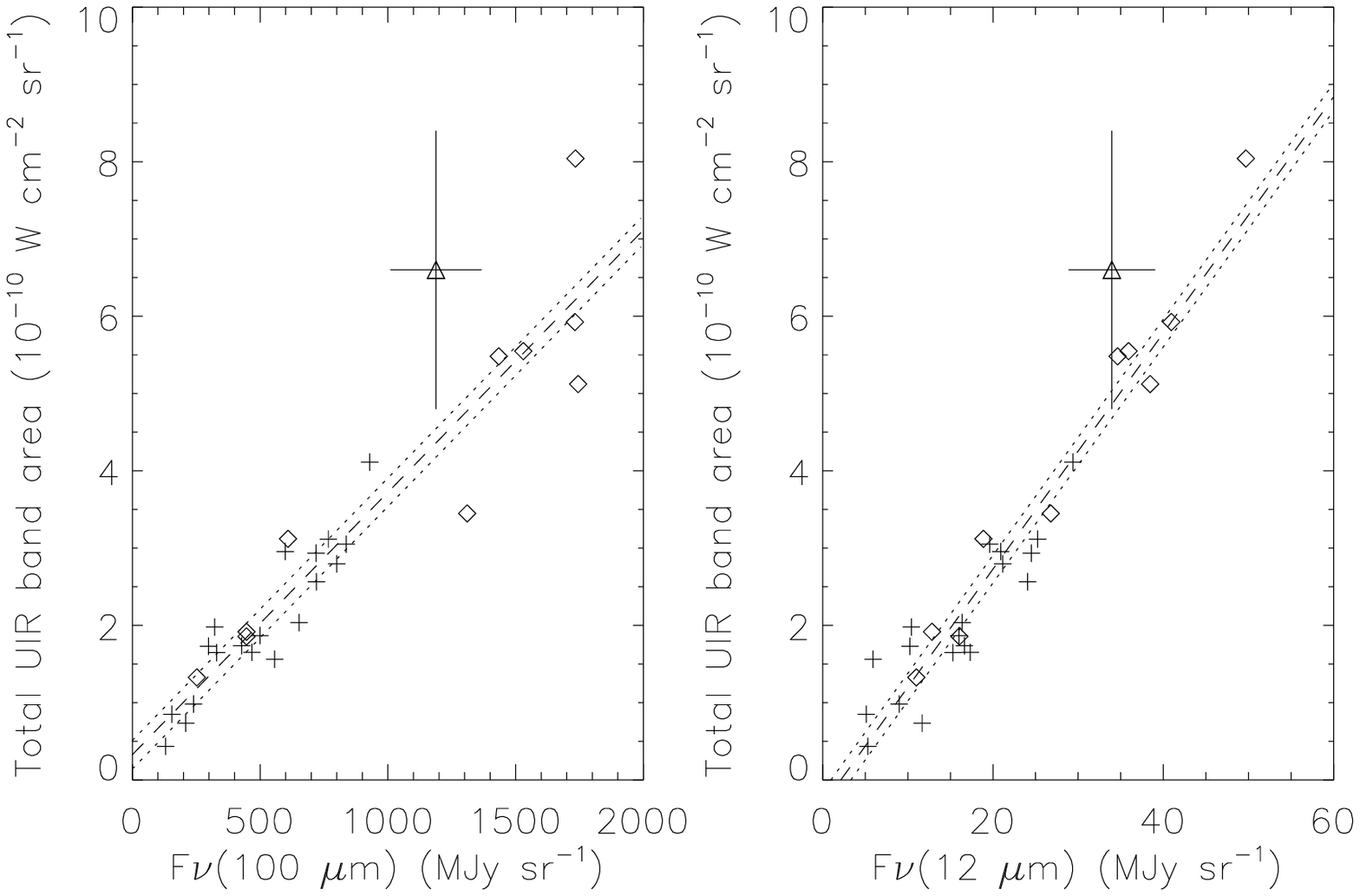}}
  \caption{
	 Total UIB flux in the 5.8-11.6 \um\ range
         versus \bold{a)} IRAS 
	 100 and \bold{b)} IRAS 12 \um\ flux from ISSA. Each 
	 panel shows also a least-squares fit to the 
	 data (dashed line) and one-sigma errors 
	 for the fit (dotted lines). Diamonds mark 
	 Galactic plane positions and crosses 
	 \galb\ = $\pm$1\degr\ values. \changed{Values for NGC 891 
	 (from Mattila et al. \cite{Mattila96})
	 are marked for comparison with a triangle with error bars.
	 }
	 \newchanged{Data for NGC 891 have been arbitrarily scaled 
	 (while preserving the ratio) to typical Milky Way values.}
	 }
  \label{fig:UIB_vs_IRAS}
\end{figure}

Figure \ref{fig:UIB_vs_IRAS} presents the total 5.8 - 11.6 \um\ UIB
emission plotted against the IRAS 100 and 12 \um\ fluxes. In both cases 
linear least-squares fits to the data and one-sigma error estimates 
are also shown.
A high correlation with the IRAS/ISSA 12 \um\ flux is expected, as the flux in
this band is dominated by UIB structures (Puget et al. \cite{Puget85}; Giard et
al. \cite{Giard89}; Mattila et al. \cite{Mattila96}) and a high correlation is
indeed found: the Pearson correlation coefficient of the IRAS 12 \um\ flux and
the total UIR band flux is 0.96. The correlation with the IRAS 100 \um\ flux
measured on the IRAS Galaxy Atlas (IGA) is equally good, with a correlation
coefficient of 0.95. A similar value (0.96) was found by Onaka et al.
\cite{Onaka96} for the correlation between UIB (7.7 \um\ band) and IRAS 100
\um\ fluxes in a larger set of observations, which sampled \changed{
only a small sector of the Galaxy ($44\degr < \gall < 54\degr$, $|\galb| \le 5\degr$)}. 
These correlation  coefficients can thus be considered
typical for the general interstellar matter.  
%
%
Previously published figures of the UIB/IRAS correlation show a deviation
towards lower UIB flux at high IRAS 100 \um\ fluxes (Tanaka et al.
\cite{Tanaka96}; Onaka et al. \cite{Onaka96}). 
\changed{This effect 
is explained by heating of large dust grains by 
intense radiation fields. Using the total FIR flux as a dust measure 
instead of using the 100 \um\ flux corrects for this effect (Onaka
\cite{Onaka00}). 
A trend towards low UIB to FIR ratios is still evident at very high 
radiation densities ($G_0 > 10^3 - 10^4\;G_{\sun}$, where 
$G_{\sun}$ is the total radiation density in the solar vicinity),
but such conditions are never seen in the general interstellar medium.}
%
Most of the UIB flux included in
our plots is emitted by the 7.7 \um\ band; 
we see no change of slope in the UIB/100 \um\ relation up to 100 \um\ 
flux of 1700 MJy/sr. 

\begin{figure}
  \resizebox{1.03 \hsize}{!}{\includegraphics{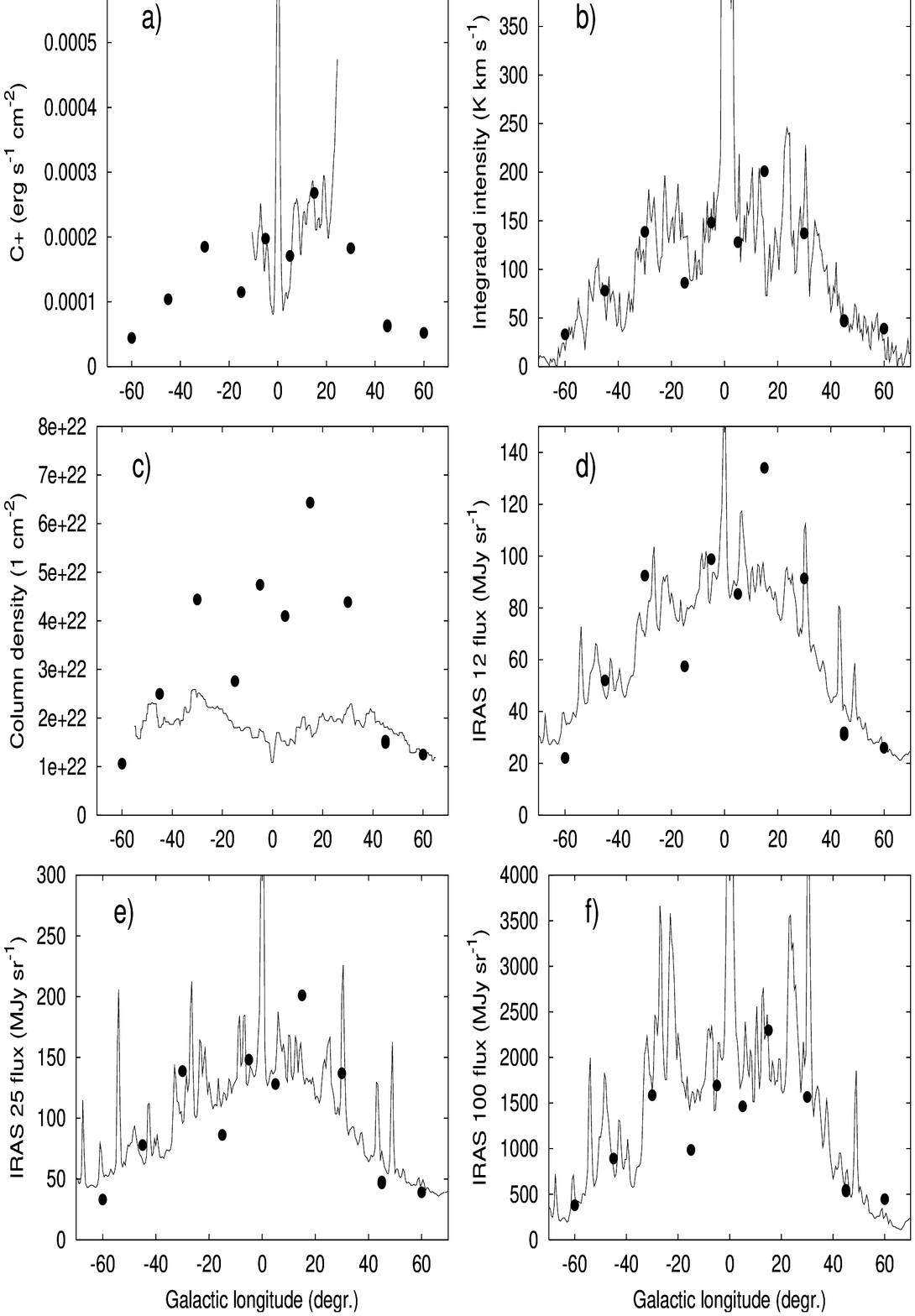}}
  \caption{
	 Comparison of UIB emission distribution with
	 \bold{a)} Diffuse gas (\ion{C}{II}, 158 \um), Nakagawa et al.
	 \cite{Nakagawa98});
	 \bold{b)} Molecular gas (\element[][12]{CO}, Dame et al. \cite{Dame01});
	 \bold{c)} Neutral hydrogen (\ion{H}{I} column density, Dickey \& Lockman
      	      \cite{Dickey90});
	 \bold{d)} IRAS 12 \um\ flux;
	 \bold{e)} IRAS 25 \um\ flux;
	 \bold{f)} IRAS 100 \um\ flux.
	 The UIB data is marked with filled circles and have
	 been scaled to match the 
	 outer parts of the comparison profile in all panels.
	 }
  \label{figure:longitudal_profiles}
\end{figure}

A comparison of the longitudinal profile of UIBs with profiles for several other
interstellar components is shown in Fig.~\ref{figure:longitudal_profiles}.
Dense gas and dust tracers (\element[][12]{CO} and IRAS far-infrared fluxes)  follow
the general UIB distribution as described in 
Sect.~\ref{section:distribution}, but the profile for atomic gas (\ion{H}{I} and
\ion{C}{II}) displays a much less concentrated profile with a central minimum.
The strong correlation with molecular gas provides a tool for
further analysis of the properties and general distribution of UIB carriers in
the Milky Way: the molecular gas tracers can be used for distance estimation.
It should be noted that the resolution of Galactic plane surveys in the radio and
infrared domains is generally less than our  field of view (1\arcmin) and thus
the details of the profiles should not be compared.

%

Our data probes the correlation between 100 \um\ and UIR band
emission \changed{in the general ISM} at two very different angular scales, \changed{
the largest scale being defined by the grid step and the smallest by the size of the 
ISOPHOT-S pixel, ie. 24\arcsec. Large-scale correlation between 100 \um\ emission 
and UIBs is proved by point-by-point comparison of the fluxes in 
Fig.~\ref{fig:UIB_vs_IRAS}.
} 
There is no obvious reason why the 100 \um\ and UIR band emission should
be so  strongly correlated at \changed{any} scales in the general ISM. 
The IRAS 100 \um\ band is dominated by thermal emission of large, cool dust grains, while 
the UIB spectrum stems from much smaller, thermally fluctuating particles. 
The conclusion is that formation and subsequent evolution 
and dynamics of the UIB carriers and large dust grains in the ISM must follow
similar lines.  The observed correlation also constrains the energy budget
of UIB carriers: the heating source for UIBs and large grains must be the same
or at least the energy sources for UIBs and FIR emission be strongly
correlated. Heating of large grains in the general ISM is dominated by the 
visible -- near infrared part of the interstellar radiation field. The same part
of the interstellar radiation field must thus be important for the 
heating of UIB carriers.
\changed{
According to Sodroski et al. (\cite{sodr97}), the relative abundance of UIB 
carriers with respect to large grains rises by a factor of two at Galactic radii larger 
than 8.5 kiloparsecs. This change should be seen in the UIB to 100 \um\
ratio. Our dataset samples mainly the inner Milky Way: even the \galg$\pm$60
lines of sight pass within 7.5 kpc from the Galactic center. Only a minor 
contribution in each spectrum is expected to arise from the outer parts 
of the Galaxy and thus this dataset cannot be used to limit the variation 
of the UIB to 100 \um\ ratio at Galactic radii larger than 8.5 kpc.}
%
%
%

%
%
%
\subsection{\object{NGC 891} and the Milky Way }\label{sec:891}
\begin{figure}
  \centering 
  \resizebox{\hsize}{!}{\includegraphics{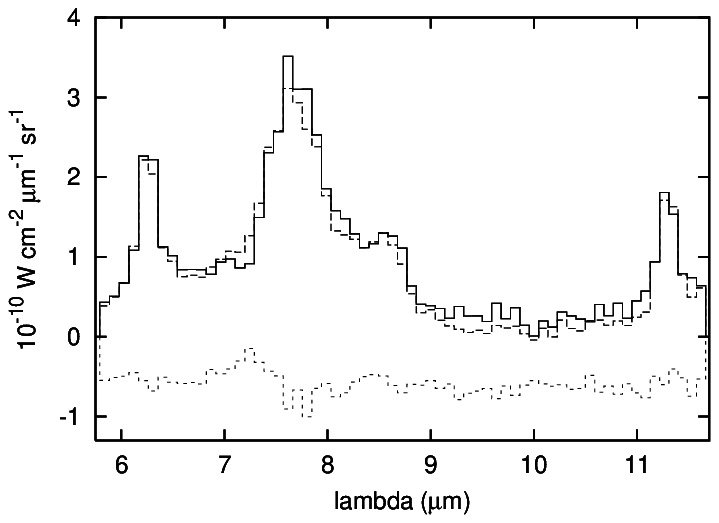}}
  \caption{
	 Comparison of the UIB spectra of \object{NGC 891} and the Milky Way.
	 Solid line: Average of Milky Way spectra. 
	 Dashed line: Average of central region of \object{NGC 891} 
	 (see Mattila et al. \cite{Mattila99}). 
	 Dotted line: The difference of the spectra mentioned above, 
	 shifted down by \mbox{0.5 \texttimes\ \punit}.
	 }
  \label{fig:ngc891}
\end{figure}

The UIB emission of the edge-on spiral galaxy \object{NGC 891} has recently been mapped
by Mattila et al. (\cite{Mattila99}). This study used the same instrument and a
similar  observing scheme as our Milky Way study. \object{NGC 891} is in many ways
similar  to our own Galaxy and this similarity is also evident in the UIB
spectrum.   An average spectrum of \object{NGC 891} was produced by summing all observed
spectra within 144\arcsec\ from the  center of the galaxy. This average, the
corresponding average spectrum  of the Milky Way and their difference  is shown
in Fig.~\ref{fig:ngc891}. The shapes of the UIB spectra in the  6-12 \um\ range are
\changed{almost} identical. 
This is remarkable especially  since the two datasets sample the ISM of the
target galaxies in different ways: Our Milky Way dataset targets small
fields in the general  ISM. In \object{NGC 891} the ISOPHOT pixel size (24\arcsec)
corresponds to 1 kpc and  there is no way to separate the contribution from
giant molecular clouds or point sources. 
\changed{A survey of UIB spectra of 
various types of spiral galaxies (Helou et al. \cite{Helou00}) shows that 
the shape we see in Fig.~\ref{fig:ngc891} 
is typical not only for Milky Way-type spirals, but for spirals in general.}

\changed{
The only, tentative, difference between the spectra of NGC 891 and the Milky Way  is the
presence of a small excess feature at 7.1 \um\ in the difference spectrum of NGC 891
minus the Milky Way average. This bump is also present in the average
UIB spectrum of 28 spiral galaxies (Helou et al. \cite{Helou00}) and less clearly in the ISO SWS 
spectrum of the Galactic center (Lutz et al. \cite{Lutz96}). 
A recent summary of UIR band data (Peeters et al. \cite{peeters02}) reveals that 
this feature is rather strong in such objects as \object{HD 44179} (the central star of 
the "Red Rectangle") while the spectra of low-radiation-density objects lack this band.
The presence of the 7.1 \um\ bump in the average spectrum of NGC 891 
reveals a significant and expected contribution from compact sources, 
\emph{if} this feature truly is missing from the spectrum of the diffuse ISM.}

The peak and average \emph{observed} surface brightness of UIR bands are
identical in \object{NGC 891} and the plane of the Milky Way (Fig.~\ref{fig:ngc891}). 
This \changed{must be a coincidence} since the viewing geometry and the beam filling 
factor are not the same for these two targets. 
In the Milky Way the UIB emission uniformly fills~ the small (24\arcsec) detector beam 
and a single  beam includes sources at both small and large distances. In the case of
\object{NGC 891}  the distance to the constituent sources is constant, but the real
surface brightness varies across a single ISOPHOT-S detector pixel (see Fig.~2 in
Mattila et al. \cite{Mattila99}). 

\changed{The ratios of the total UIB flux to the 100 \um\ and 12 \um\ fluxes of NGC 891
are similar to the ratios for individual lines-of-sight in the  Milky Way.  Data for NGC
891 are included in Fig.~\ref{fig:UIB_vs_IRAS} and marked with triangles with error bars.
Both UIB/12 \um\ and UIB/100 \um\ ratios for NGC 891 lie above typical Milky Way values by
30-40\%.  However, the uncertainty of the ratios ($\pm$30\%) is such that the difference
is not significant. Not only the shape of the UIB spectrum  but also the total
intensity compared with other dust components is similar in the  Milky Way and in NGC 891.
(Note that the units for NGC 891 and the Milky Way data are different, surface brightness
versus total flux, and only the ratios can be  meaningfully compared here). The flux values
for NGC 891 shown in  Fig.~\ref{fig:UIB_vs_IRAS} were arbitrarily scaled -- while
preserving the ratio --  to fit into the brightness range defined by Galactic values.}

In spite of the different sampling of the ISM in the Milky Way and \object{NGC 891}  by
our PHT-S observations the UIB spectra are almost identical. Therefore, the
spectrum in  Fig.~\ref{fig:ngc891} can be used as a template spectrum for the
6-12 \um\ emission of large, late-type spiral galaxies in general until new IR
observatories in space  enlarge the available data base. 

%
\subsection{Band ratios in the PAH model}
The observed band ratios can be converted into physical properties of  the
emitting particles. This conversion  is naturally model-dependent as the
radiative properties of very small, solid  interstellar particles are still
poorly understood. We adopt the PAH model  of Draine \& Li (\cite{Draine01})
for our analysis. The observed mean values for  band ratios (11.3/7.7 = 0.27,
6.2/7.7 = 0.21) correspond to a somewhat ionised PAH mixture with a  typical
particle size around 200 carbon atoms heated by an average  interstellar
radiation field. Draine \& Li define band ratios in the same way as we do, 
ie. as the ratio of areas of Cauchy-profile fits to the bands.
Their model results are directly comparable with the ratios used in this paper. 
Previous papers on UIR bands have frequently defined the band area
as the integrated flux above an estimated baseline. Ratios derived from 
such estimates are not directly comparable with our results or the model
values in Draine \& Li since a large part of the intensity in Cauchy profiles 
lie outside the central peak.

While each of our observations samples  several diffuse 
and molecular clouds,
it is to be expected that the conditions in these clouds do not vary wildly and
so average values over observed lines-of-sight still represent typical 
values in  the general ISM. 

%



\subsection{Absorption effects and the 9-11 $\mu$m continuum}
\label{sec:extmodel}
The study of mid- and far-IR continuum emission of low 
brightness sources is strongly
affected by the zodiacal foreground. We have tried to minimize this
error source by using two OFF-position measurements for each longitude. The
zodiacal-subtracted spectra include only a  trivial continuum component; the
remaining  flux in the  9-11 \um\ range is a product of the wide wings of the
7.7 \um\ and 11.3 \um\ bands.
\newchanged{The lack of flux in this interval} 
could be attributed to a number of
phenomena. The most simple is a true lack of emission in this wavelength range 
from the diffuse ISM. \newchanged{A weak continuum component (less 
than our detection limit) was reported
by the Onaka et al. (\cite{Onaka96}). The Mid-IR Spectrometer (MIRS) 
of IRTS had a much larger fields-of-view than ISOPHOT-S and thus a 
contribution from compact UIB emitters within the field of view 
is probably present in the published MIRS spectra of the diffuse ISM. 
}

Another possible cause of the low flux is the broad silicate absorption feature
centered at 9.7 \um.  This structure has been observed in the diffuse medium
towards the Galactic  center (Rieke \& Lebowsky \cite{Rieke85}) and several
local  sources (Knacke \& Gaustad \cite{Knacke69}; Gillett et al.
\cite{Gillett75})  as well as in nearby spiral galaxies (Glass et al.
\cite{Glass82};  Imanishi \& Ueno \cite{Imanishi00}). 

We can solve the attenuation caused by the silicate band if we assume
spatially constant values for the emissivity ($\epsilon_{\lambda}$) and the
absorption coefficient ($\kappa_{\lambda}$).  In the unobscured case, the surface
brightness is simply $I_0 = \epsilon_{\lambda} r$, where $r$ is the path length. If 
$\kappa_{\lambda}$ is  $> 0$, the intensity is  
\begin{equation} I =
\frac{\epsilon_{\lambda}}{\kappa_{\lambda}}(1-\mathrm{e}^{-\tau_{\lambda}}) =
\frac{I_0}{\tau_{\lambda}}(1-\mathrm{e}^{-\tau_{\lambda}})  .
\label{eq:ext}
\end{equation}
%
Optical thickness at the center of the silicate feature can be estimated
from A$_V$: 
$\tau_{Si}$ is 0.05 - 0.07 \texttimes\ A$_V$
(Aitken \cite{Aitken81}; Rieke \& Lebowsky \cite{Rieke85}; van der Hucht et
al. \cite{vdHucht96}); we adopt an average value, 0.052, as calculated in the
summary of observations in  Mathis (\cite{Mathis98}).  Resulting loss of
intensity near the 9.7 \um\ minimum is $I/I_0 = 0.8-0.5 \; (\tau_{Si} =
0.5-1.4)$, as the average optical  attenuation is $A_V \approx 1
\mathrm{^m kpc^{-1}}$ and a typical path length within the Milky Way is 10-20
kiloparsecs. The total loss of flux at the center of the 9.7 \um\
silicate  absorption feature is thus only 20-50\% and the average  \emph{unobscured}
flux in the 9-11 \um\ minimum would be less than \mbox{0.4 \texttimes\ \punit} or
$<15$\% of the peak flux in the 7.7 \um\ band. It is not surprising that the
general ISM displays such a faint continuum  component as a steep continuum is
typical for sources with high UV energy  density and a hard UV spectrum; 
moderate sources such as reflection  nebula NGC 7023
display only a trace of continuum emission at wavelengths shorter than 20 \um\
(Moutou et al. \cite{Moutou98}).

\begin{table}
   \caption[]{Effects of extinction in UIR band ratios.}
   \label{tab:ext}
   \[
\begin{tabular}{r r l l l}
   \hline
   \hline
   & & \multicolumn{3}{c}{Band ratios} \\
   $\tau_{\mathrm{Si}}$ & A$_V$ & 6.2/7.7 & 8.6/7.7 & 11.3/7.7\\
   \hline
   $\to$ 0.0 & $\to$ 0.0 & 0.24 & 0.17 & 0.30 \\
         0.1 &       1.9 & 0.24 & 0.17 & 0.29 \\
         0.5 &       9.6 & 0.24 & 0.16 & 0.28 \\
         1.0 &       19  & 0.24 & 0.14  & 0.26 \\
         5.0 &       95  & 0.26 & 0.049 & 0.17 \\
    $\infty$ & $\infty$  & 0.27 & 0.017 & 0.12 \\
   \hline
   \end{tabular}
   \]
\end{table}

This analysis can be taken further by using a real extinction law and 
calculating the change of band profiles with increasing optical depth.  
The relation between  UIB emitters and absorbing particles must then be defined.
The high correlation between  large dust grains and UIR band strength leads to
the conclusion that  these populations are well mixed: 
$\kappa_{\lambda} \propto \epsilon_{\lambda}$.  
In this case it is not possible to talk about the intensity of unobscured
profile as internal extinction will always be present in any UIB-emitting 
region. The shape of the UIB profile  can still be defined as the limiting
values when  $\tau \to 0$.
With these assumptions it is possible to model changes in
profile shapes by using the first form of Eq.~(\ref{eq:ext})
individually for each wavelength of the spectrum. 
We adopted $\tau_{Si}$ as the measure of extinction and
calculated optical thickness for other wavelengths from 
the silicate extinction law in Draine \& Li (\cite{Draine01}). 
Finally, the shape of the unobscured spectrum was approximated 
with the fit to the average of observed spectra,
corrected using Eq.~(\ref{eq:ext}) and $\tau(Si) = 1$. 

\begin{figure}
  \resizebox{\hsize}{!}{\includegraphics{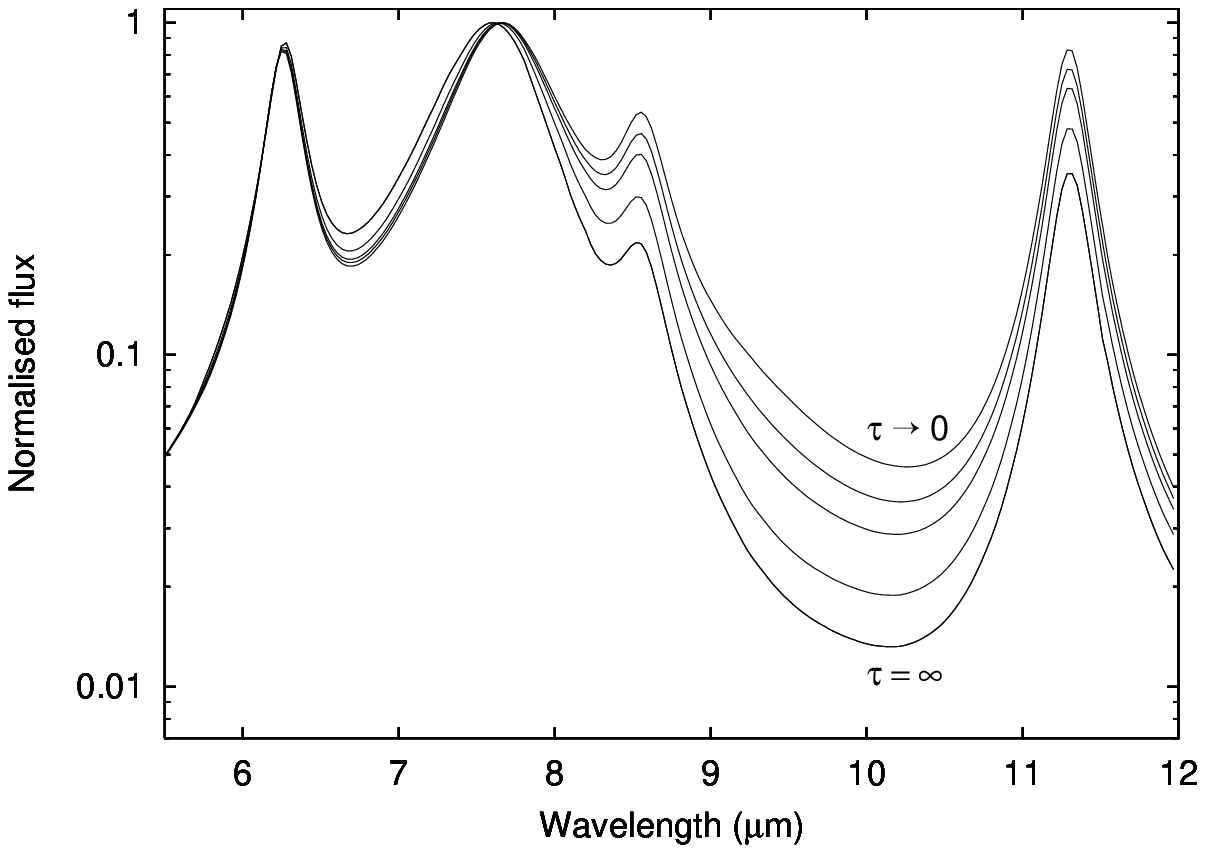}}
  \caption{
	 Absorptions effects in UIR band profiles. Profiles were calculated 
	 using the model described in Sect.~\ref{sec:extmodel}
	 for (from top to bottom) $\tau(9.7 \um)$ = 
	 0, 1, 2, 5 and $\infty$. Each spectrum is normalized to 1.0 
	 at the top of the 7.7 \um\ band.
	 }
  \label{fig:ext}
\end{figure}

The resulting observable spectra  
for $\tau$ = 0, 1, 2, 5 and $\infty$ are shown in Fig.~\ref{fig:ext}.
Each spectrum is normalised to unity at the peak of the 7.7 \um\ band. 
\changed{ From the first form of Eq.~(\ref{eq:ext}) we see that when $\tau =
\infty$ the observed intensity is simply  $\epsilon_{\lambda} /
\kappa_{\lambda}$.}
Model band ratios were calculated from the modeled profiles.
Table~\ref{tab:ext} lists these band ratios for a range of  values of
$\tau_{Si}$.  The first \changed{row} shows the band ratios in the unobscured
spectrum and the following \changed{rows} the observable ratios for $\tau_{Si}$ = 0.1,
0.5, 1, 5 and  $\infty$.
It is clear that the interstellar extinction is a minor contributor to the
observed band ratios and  band shapes in the diffuse ISM. The shape of the
minimum around 10 \um\ is also practically unaffected by extinction in this
simple test  calculation.
The changes become apparent only when optical
thickness  at the center of the silicate feature becomes larger than unity. In
such conditions the 8.6 \um\ band is strongly damped and the 9--11 \um\
minimum becomes deeper; the 11.3/7.7 \um\ band
ratio is less affected. The 6.2/7.7 \um\ band ratio does not change with
increasing extinction because the extinction curve is flat in this wavelength
range.

\section{Conclusions} 
We have proven that the UIB emission signature of the general ISM  is constant
in a sample of lines-of-sight covering the inner Milky Way. This independence
is indeed what one expects to see if the UIBs in the  diffuse component are
mainly heated by single-photon events. The observed shape and calculated band
ratios are typical for UIB sources with low particle density and a low UV 
flux. \changed{No interstellar continuum emission component is seen in the
observed wavelength range.}
\changed{ The UIB spectrum of other spiral galaxies differ only slightly from
the  Milky Way spectrum and the spectrum presented in Fig.~\ref{fig:ngc891} 
can be used as a template for the  6-12 \um\ emission of spirals. A weak
emission bump around 7.1 \um,  present in the total spectrum of spirals, seems
to be lacking in the  spectrum of the diffuse ISM. The significance of the
observed variability of this  feature among galactic objects and galaxies is still unknown.}

The distribution of UIB emission in the Galactic ISM follows the general
distribution of large dust grains (IRAS 100 \um\ emission) and
molecular (CO) gas. The same tendency is seen in the high correlation between 
IRAS 100 \um\ and UIB emission \changed{at large (10\degr)} angular scales.
Such a strong  correlation links the history and energy budget of UIB carriers
directly to the large grain population: the physical processes that are
responsible for creating and  heating the two particle populations must be
closely related.

Finally, we have analyzed the effect of interstellar extinction on our 
observations and find that the silicate extinction feature 
centered at 9.7 \um\ only contributes weakly to lowering the continuum 
at 10 \um\ if silicates are well mixed with UIB carriers.
%
%
%
The extinction effects on band
ratios are also found to be negligible as long as the silicon absorption 
feature is not optically thick.

%
\begin{acknowledgements}
Contributing ISOPHOT Consortium institutes are DSRI, 
DIAS, RAL, AIP, MPIK, and MPIA.
ISOPHOT and the Data Centre at MPIA, Heidelberg, are funded by the Deutsches
Zentrum f\"ur Luft- und Raumfahrt and the Max-Planck-Gesellschaft.

The ISOPHOT data presented in this paper were reduced using PIA, which is a 
joint development by the ESA Astrophysics Division and the ISOPHOT 
Consortium with the collaboration of the Infrared Processing and 
Analysis Center (IPAC). Contributing ISOPHOT Consortium institutes are 
DIAS, RAL, AIP, MPIK, and MPIA.

The ISO Spectral Analysis Package (ISAP) is a joint development by the 
LWS and SWS Instrument Teams and Data Centers. 
Contributing institutes are CESR, IAS, IPAC, MPE, RAL and SRON.

This study was supported by the Academy of Finland Grants no. 173727 and 174854
and by the Finnish Graduate School for Astronomy and Space Physics.
\end{acknowledgements}

\end{document}